\documentclass[twocolumn,amsmath,amssymb,prb,floatfix]{revtex4}
\usepackage{psfig}
\makeatletter
\usepackage{graphicx}
\begin{document}

\title{Electronic structure, exchange interactions and Curie temperature in diluted III-V
magnetic semiconductors: (GaCr)As, (GaMn)As, (GaFe)As.}

\author{
L.M. Sandratskii\thanks{lsandr@mpi-halle.de} and P. Bruno}

\affiliation{Max-Planck Institut f\"ur Mikrostrukturphysik, D-06120 Halle, Germany }
\begin{abstract}
We complete our earlier (Phys. Rev. B, {\bf 66}, 134435 (2002)) study of the electronic structure, 
exchange interactions and Curie temperature in (GaMn)As and extend the study to two other
diluted magnetic semiconductors (GaCr)As and (GaFe)As. Four concentrations of the 
3d impurities are studied: 25\%, 12.5\%, 6.25\%, 3.125\%. 
(GaCr)As and (GaMn)As are found to possess a number of similar features.
Both are semi-metallic and ferromagnetic, with similar properties of the
interatomic exchange interactions and the same scale of the Curie temperature.
In both systems the presence of the charge carriers is crucial for establishing 
the ferromagnetic order. An important difference between two systems is in the 
character of the dependence on the variation of the number of carriers. The 
ferromagnetism in (GaMn)As is found to be very sensitive to the presence of the donor defects,
like As$_{\rm Ga}$ antisites. On the other hand, the Curie temperature of (GaCr)As
depends rather weakly on the presence of this type of defects but decreases strongly
with decreasing number of electrons.
We find the exchange interactions between 3d atoms that
make a major contribution into the ferromagnetism of
(GaCr)As and (GaMn)As and propose 
an exchange path responsible for these interactions. The properties of (GaFe)As are
found to differ
crucially from the properties of (GaCr)As and (GaMn)As. (GaFe)As does not show a trend to
ferromagnetism and is not half-metallic that makes this system unsuitable for 
the use in spintronic semiconductor devices.   
\end{abstract}
\maketitle

\section{Introduction}

The discovery of the ferromagnetism in (GaMn)As \cite{ohno_gaas} with the Curie temperature 
of Ga$_{0.947}$Mn$_{0.053}$As as high as 110 K
attracted much attention to the III-V diluted 
magnetic semiconductors. This attention was stimulated by both the possibility of deeper
understanding of the fundamentals of long-range ferromagnetism in semiconductors
and the practical need for the 
ferromagnetic diluted magnetic semiconductor (DMS) materials with high Curie temperature 
for the realization of the semiconductor spin-electronic devices.   

On the theoretical side both the model-Hamiltonian approach
(see recent reviews \cite{dietl_rev,dms_mac}) 
and the parameter-free
calculations within the density functional theory (see, e.g., a review \cite{sanvito_rev})
are intensively used to study the magnetism of the DMS. 
One of the important  
directions of the both theoretical techniques is the calculation of the 
exchange interactions and Curie temperature in the DMS systems (see, e.g.,
\cite{diofma,mark01,bopa,jukosi,erpe,sabr02_1}).  

In the previous paper \cite{sabr02_1} (hereafter referred to as I) we used the 
supercell and frozen-magnon approaches to study the
exchange interactions and Curie temperature of (GaMn)As for four different Mn concentrations. 
The present work extends the study 
to the cases of Cr and Fe impurities. 
Since a Cr atom has one electron less
and an Fe atom one electron more than a Mn atom, comparison of the
results for three DMS systems allows the investigation of 
the trends in the variation of magnetic properties within the series of the DMS.

One of the issues attracting much attention is the 
dependence of the 
Curie temperature on the presence of nonmagnetic defects. 
Such defects can be introduced
purposely, e.g., by co-doping, or appear in an uncontrolled manner
during the sample preparation
(e.g., As$_{\rm Ga}$ antisites). Since these defects change the density of carriers 
they can strongly influence the magnetism of the DMS systems. \cite{dietl_rev,dms_mac,sanvito_rev} 
This influence is also addressed in the paper.

The Cr and Fe impurities in GaAs were recently studied theoretically. \cite{mark01,sato_rev}
In Ref.  \cite{mark01} the exchange interactions within small clusters of the 
3d impurities have been studied. In Ref. \cite{sato_rev}, the coherent potential
approximation has been used to estimate the energy difference between the ferromagnetic
and spin-glass states. In both works the authors came to the conclusion of the strong
ferromagnetic exchange interactions in (GaCr)As. No trend to the ferromagnetism has been
found in (GaFe)As. First attempts to synthesize these systems have been reported. 
\cite{sazaak,mokata}
Although the dominant magnetic interactions between Cr atoms were found to be ferromagnetic
no high Curie temperature in (GaCr)As was detected. \cite{sazaak} Interestingly,
the zinc-blende CrAs that has recently been successfully grown on GaAs is ferromagnetic
with the Curie temperature higher than room temperature. \cite{akmash}
This system has been designed on the basis of the calculations within the framework of
the density-functional theory (DFT) and its properties are is good agreement with theoretical 
predictions. \cite{akmash} 

The purpose of the present work is to provide detailed study of the exchange interactions
and Curie temperature in the series of diluted magnetic semiconductors:
(GaCr)As, (GaMn)As, (GaFe)As. Much attention is devoted to the comparative analysis of the
systems. One of our aims is to further stimulate experimental interest to (GaCr)As and 
(GaFe)As systems. 

The remainder of the paper is organized as follows. In Sect. II we discuss a simple
two-band model studying the relation between the magnetic structure, the band occupation
and the band energy. In Sect. III we briefly present the calculational approach.
In Sect. IV we discuss the calculational results for (GaMn)As, (GaCr)As, and
(GaFe)As systems. Our conclusions are given in Sect. V.

\section{Kinetic exchange and band occupation}
\label{sect:two-band} 

It is common to treat the magnetism of the DMS in terms of the competition between 
the antiferromagnetic superexchange and ferromagnetic 
kinetic exchange through charge carriers (see, e.g., Refs. \cite{akai,mark01})
The definitions
of different types of exchange interactions rely on different model Hamiltonians
and their perturbative treatments. \cite{magnetism_anderson} 
Since the DFT is not based on a model Hamiltonian
approach and does not use a perturbative treatment, 
various exchange interactions appear in the calculational results in 
a mixed form.  
In this situation the studies of simple 
models of electron systems relevant to the problem provide information 
useful in qualitative interpretation of the DFT results.   

Here we consider a simple two-band tight-binding model of itinerant electrons  
experiencing local exchange fields of atomic magnetic moments. 
We consider helical configurations of the atomic
moments and study the dependence of the band energy of the system on the 
magnetic structure. Three cases are discussed: completely filled bands, almost
empty bands and almost filled bands.
The helical structures are defined by the formula
\begin{equation}
\label{spir}
{\bf e}_{n} = \left( \cos ({\bf q} \cdot {\bf R}_n)
 \sin \theta,\: \sin
({\bf q} \cdot {\bf R}_n) \sin \theta,
 \:\cos \theta\right)
\end{equation}
where ${\bf R}_n$ are the lattice vectors, ${\bf q}$ is the wave vector of the helix,
${\bf e}_{n}$ is the unit vectors in the direction of the magnetic moment at site
${\bf R}_n$, polar angle $\theta$ gives the deviation of the moments from the $z$ axis.
The helical structures 
allow to describe broad range of magnetic configurations
from collinear ferromagnetism ($\theta=0$ or ${\bf q=0}$) to collinear antiferromagnetism 
(${\bf q}=\frac{1}{2}{\bf K}$ and $\theta=90^\circ$, ${\bf K}$ is a reciprocal lattice vector).

The tight binding method for spiral structures was discussed
in its general form in Ref. \cite{sand86}.
By neglecting the difference in the
spatial dependence of the basis functions with opposite spin projections
and by preserving only the single-center matrix elements of the exchange
potential we arrive at the following simple form of the secular matrix:
\begin{widetext}
\begin{equation}
\label{tb-mat}
\left(\begin{array}{cc}\cos^2(\frac{\theta}{2})H_-+\sin^2(\frac{\theta}{2})H_+
-\frac{\Delta}{2}&
-\frac{1}{2}\sin\theta(H_--H_+)\\
-\frac{1}{2}\sin\theta(H_--H_+)&
\sin^2(\frac{\theta}{2})H_-+\cos^2(\frac{\theta}{2})H_++\frac{\Delta}{2}
\end{array}\right)
\end{equation}
\end{widetext}
where $H_-=H({\bf k}-\frac{1}{2}{\bf q})$, 
$H_+ =H({\bf k}+\frac{1}{2}{\bf q})$, and $H({\bf k})$
describes spin-degenerate bands of a non-magnetic
crystal. 
The derivation of Eq. (\ref{tb-mat}) is based on the special symmetry of 
the helical structures.\cite{adv}
This model describes one electron band with two opposite spin projections.
At each lattice site ${\bf R}_n$ the band-electron states experience a local
exchange field of strength $\Delta$. 
The model neglects the effects of hybridization between magnetic
and band electrons and therefore is not suitable for the description of the 
superexchange.\cite{magnetism_anderson} 
It, however, takes into account interactions characteristic 
for the kinetic exchange through charge carriers.\cite{magnetism_anderson,deGennes,mrsafr,comm2} 
We will study the dependence of the 
energy of the band electrons on the magnetic configuration. 

\begin{figure}
\caption{The energy bands (thick solid line) of the spiral structure with ${\bf q}=0.4$, 
$\Delta=2$, 
and $\theta=45^\circ$. $H({\bf k})=1-\cos(\pi{\bf k})$. The thin solid lines show the
bands of a ferromagnetic configuration. The broken lines give the ferromagnetic bands
shifted in the reciprocal space according to the given {\bf q} (see the text). The minimal
energy of the ferromagnetic configuration is lower than for the spiral. On the other
hand, the maximal energy of the ferromagnet is higher than for the spiral.
\label{fig:spiral_bands}}
\centerline
{\includegraphics[width=8cm,angle=-90]{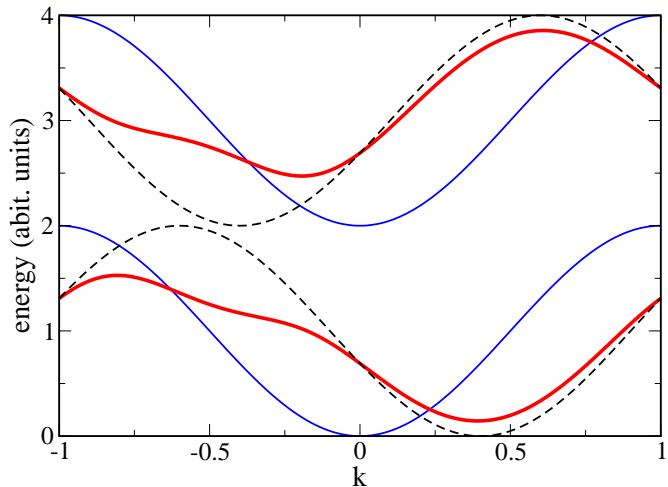}}
\end{figure}

The eigenvalues of the matrix (\ref{tb-mat}) have the form
\begin{eqnarray}
\label{tb-eigv}
\lefteqn{\varepsilon_{\pm}({\bf k})=
\frac{1}{2}(H_-+H_+)}\\
&&\pm(\frac{1}{4}(H_--H_+)^2-
\frac{1}{2}\Delta\cos\theta(H_--H_+)+\frac{1}{4}\Delta^2)^\frac{1}{2}\nonumber
\end{eqnarray}
and are illustrated in Fig. \ref{fig:spiral_bands}. 
To understand the structure
of the electron bands of the spiral it is important to notice that the ferromagnetic
configuration of the local moments can be considered as a spiral with
$\theta=0$ and arbitrary ${\bf q}$. The consideration of the ferromagnet as 
a spiral leads to a rigid shift of the spin-up subband to the right by $\frac{{\bf q}}{2}$
and the shift of the spin-down subband to the left by $\frac{{\bf q}}{2}$. If now
the angle $\theta$ becomes nonzero the shifted bands of the opposite spin hybridize 
because of non-zero off-diagonal elements of the secular matrix [Eq. (\ref{tb-mat})].
This leads to the repulsion of the states 
that are close in energy
(e.g., the states at ${\bf q}=-0.5$ in Fig. \ref{fig:spiral_bands}). Additionally,
there is an intraband mixing of the states separated by vector ${\bf q}$ in the
reciprocal space (see the diagonal elements of the secular matrix.

First, let us consider the case of completely filled bands. 
In this case the total energy equals to
\begin{equation}
\label{eq:etot_full_bands}
E_b=\int_{\rm BZ} d{\bf k}[(\varepsilon_{-}({\bf k})+\varepsilon_{+}({\bf k})]=
2\int_{\rm BZ} d{\bf k}H({\bf k})  
\end{equation}
and does not depend on the magnetic configuration. 
(In Eq. (\ref{eq:etot_full_bands}) the integration is carried out
over the first Brillouin zone (BZ).) This means that the
kinetic exchange taken into account by Eq.(\ref{tb-mat}) does not influence the magnetic
configuration in the system with filled bands. 
The superexchange not accounted for by Eq.(\ref{tb-mat})
plays the main role in the case of completely filled bands.

Now let us consider almost empty bands. In this case the 
minimum of the electron energy 
corresponds to the ferromagnetic structure. To prove this we 
show that any deviation from the ferromagnetic structure increases the
the minimal energy of the band states (see Fig. \ref{fig:spiral_bands} for
an illustration).

It is sufficient to consider the lower branch of the eigenvalues $\varepsilon_{-}({\bf k})$.
Let us take an arbitrary {\bf k}. If
If  $H_-({\bf k})<H_+({\bf k})$ than 
$\varepsilon_-({\bf k})>H_-({\bf k})-\frac{1}{2}\Delta$. 
If  $H_-({\bf k})>H_+({\bf k})$ than 
$\varepsilon_-({\bf k})>H_+({\bf k})-\frac{1}{2}\Delta$.
%
Therefore, in both cases the eigenstate of the helix $\varepsilon_-({\bf k})$ is higher 
than one of the states of the ferromagnetic crystal. 
For the ${\bf k}$ vector satisfying $H_-({\bf k})=H_+({\bf k})=h$,
$\varepsilon_-({\bf k})=h-\frac{1}{2}\Delta$. 
In the usual
case of $H({\bf k})$ having minimum at the center or the boundary of the 
BZ,  $h-\frac{1}{2}\Delta$ is not the minimal energy of the ferromagnetic 
structure.
%

Also for almost filled bands the configuration of lowest energy is ferromagnetic
since any deviation
from the ferromagnetic state increases the energy of the electrons. 
To show this we first notice that the 
maximal energy of the band states always decreases with deviation from ferromagnetism.
The proof of this is similar to that given above for the minimal energy.
Since the energy of the completely filled bands does not depend on the magnetic
configuration, the highest energy of the hole states
means the lowest energy of the system.

In the case of exchange splitting $\Delta$ much bigger than the width of the
nonmagnetic band $H({\bf k})$ the low-energy eigenvalue of the secular matrix given 
by Eq.(\ref{tb-mat}) takes the form
\begin{eqnarray}
\label{eq:subband}
\varepsilon_{-}({\bf k})&=&\cos^2(\frac{\theta}{2})[H({\bf k}-\frac{1}{2}{\bf q})-\frac{\Delta}{2}]\\
&+&\sin^2(\frac{\theta}{2})[H({\bf k}+\frac{1}{2}{\bf q})-\frac{\Delta}{2}]\nonumber
\end{eqnarray} 
It can be easily verified that this subband fulfills all three properties 
formulated above for the set of two bands.
 
These properties of the model agree qualitatively with those obtained within other
models, e.g., within the model of two interacting impurities. \cite{mark01}
We will refer to these properties in the discussion of the
results of the DFT calculations in the following sections. 

\section{Calculational approach}
\label{sec_calc_app}
In the calculations we use the scheme discussed in I. This scheme is
based on the supersell approach  where one of the 
Ga atoms in a supercell of zinc-blende GaAs is replaced by the 3d atom.
The calculations are performed for four values of the concentration $x$:
0.25, 0.125, 0.0625, and 0.03125.

The calculations were carried out with the augmented spherical waves \cite{wikuge} (ASW) 
method within the local density approximation (LDA)
to the DFT. In all calculations the lattice parameter 
was chosen to be equal to the experimental lattice parameter of GaAs.
Two empty spheres per formula unit have been used in the calculations.
The positions of empty spheres are (0.5, 0.5, 0.5) and (0.75, 0.75, 0.75).
Radii of all atomic spheres were chosen to be equal. 
Depending on the concentration of Mn, the super cell is cubic
(x=25\%, 
$a\times a\times a$, and x=3.125\%, $2a\times 2a\times 2a$) 
or tetragonal 
(x=12.5\%, 
$a\times a \times 2a$ and 6.25\%, $2a\times 2a\times a$).

The densities of states (DOS) presented in the paper are calculated
for self-consistently determined ferromagnetic states of the 
corresponding systems.

To describe the exchange interactions in the system
we use an effective Heisenberg
Hamiltonian of classical spins
\begin{equation}
\label{eq:hamiltonian}
H_{eff}=-\sum_{i\ne j} J_{ij} {\bf e}_i\cdot {\bf e}_j
\end{equation}
where $J_{ij}$ is an exchange interaction between two 3d sites $(i,j)$
and ${\bf e}_i$ is the unit vector pointing in the direction 
of the magnetic moments at site $i$. 

To estimate the parameters of the Mn-Mn exchange interaction we
performed calculation for the following frozen-magnon configurations:
\begin{equation}
\theta_i=const, \:\: \phi_i={\bf q \cdot R}_i
\end{equation}
where $\theta_i$ and $\phi_i$ are the polar and azimuthal angles of vector
${\bf e}_i$, ${\bf R}_i$ is the position of the $i$th Mn atom. 
The directions of the induced moments in the atomic spheres of Ga and As
and in the empty spheres were kept to be parallel to the $z$ axis. 

It can be shown that within the Heisenberg model~(\ref{eq:hamiltonian})
the energy of such configurations can be represented in the form
\begin{equation}
\label{eq:e_of_q}
E(\theta,{\bf q})=E_0(\theta)-\theta^2 J({\bf q})
\end{equation}
where $E_0$ does not depend on {\bf q} and $J({\bf q})$ is the 
Fourier transform of the parameters of the exchange interaction between 
pairs of Mn atoms:
\begin{equation}
\label{eq:J_q}
J({\bf q})=\sum_{j\ne0} J_{0j}\:\exp(i{\bf q\cdot R_{0j}}).
\end{equation}
In Eq. (\ref{eq:e_of_q}) angle $\theta$ is assumed to be small.
Using $J({\bf q})$ one can estimate the energies of the spin-wave 
excitations:
\begin{equation}
\omega({\bf q})=\frac{4}{M} [J({\bf 0})-J({\bf q})]=
\frac{2}{M(1-\cos\theta)}(E(\theta,{\bf q})-E(\theta,{\bf 0}))
\end{equation} 
where $M$ is the atomic moment of the Mn atom.
Performing back Fourier transformation we obtain the parameters of 
the exchange interaction between Mn atoms:
\begin{equation}
\label{eq:J_0j}
J_{0j}=\frac{1}{N}\sum_{\bf q} \exp(-i{\bf q\cdot R_{0j}})J({\bf q}).
\end{equation}
The calculation of $E(\theta,{\bf q})$ for different Mn concentrations
has been performed   
for uniform meshes in the first BZ. Angle $\theta$ is selected in the
proportionality region between $[E(\theta,{\bf q})-E(\theta,{\bf 0})]$
and $(1-\cos\theta)$.

The Curie temperature was estimated in the mean-field (MF) approximation
\begin{equation}
\label{eq:Tc_MF}
k_BT_C^{MF}=\frac{2}{3}\sum_{j\ne0}J_{0j}
\end{equation}

We use rigid band 
approach to calculate the exchange parameters and Curie temperature
for different electron occupations in each of the systems studied. 
This allows to simulate the influence of the As antisites and the 
nonmagnetic co-doping 
for a given concentration of 3d impurities.
We assume that the electron structure calculated for a DMS with a given
concentration of the 3d impurity is basically preserved in the
presence of defects. The main difference
is in the occupation of the bands and, respectively, 
in the position of the Fermi level.
For each electron occupation we calculated the energy of the frozen-magnon states
and the interatomic exchange parameters.

\section{Calculational results}
\subsection{(GaMn)As}
  
The main body of the calculational results for (GaMn)As is presented in I.
Here we discuss the dependence of the 
calculated Curie temperature 
on the number of carriers. The experimental studies show that the concentration 
of holes in (Ga,Mn)As
is lower than the concentration of the Mn atoms. \cite{ohno_jmmm} 
One of the important factors leading to the low concentration of holes is
the presence of the As antisites.
(See Refs. \cite{akai,sanv_apl01,korz02,sabr02_1} for earlier DFT studies of the 
influence of the As antisites on the properties of the DMS.) 

\begin{figure}
\caption{$T_C^{MF}$ of (Ga,Mn)As with different Mn concentrations
as a function of the electron number $n$.
$n=0$ corresponds to the system  Ga$_{1-x}$Mn$_{x}$As with no
additional donor or acceptor defects.  
\label{fig:T_C_of_n_Mn}}
{\includegraphics[width=8cm,angle=-90]{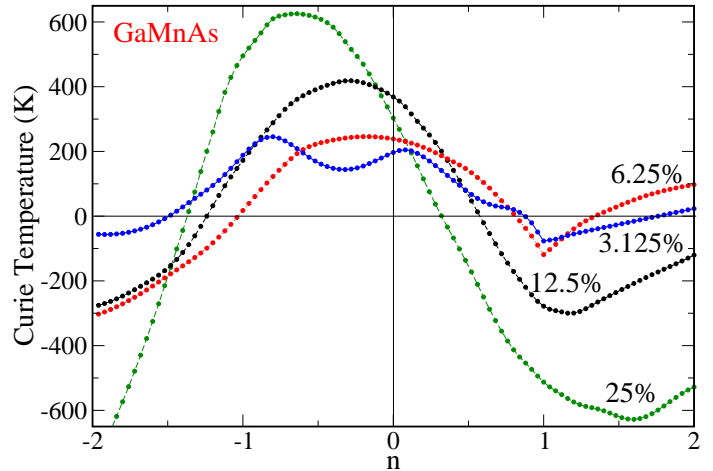}}
\end{figure}

Since As has two more valence electrons compared with Ga, each As antisite
compensate the holes produced by two Mn atoms. 
We study the influence of the 
number of electrons in the broad interval from $n=-2$ (two electrons per supercell
less) to $n=2$ (to electrons per supercell more). 
Concretely, for each frozen-magnon configuration studied we perform the 
total energy calculation for different $n$ and then for each $n$ the procedure
of the evaluation of the exchange interactions 
described in Sec. \ref{sec_calc_app} is used.
Negative values of the Curie temperature in Fig.\ref{fig:T_C_of_n_Mn}
indicate an instability of the ferromagnetic state due to dominating
antiferromagnetic interactions.

The results of the calculations are shown \cite{comm1} in Fig. \ref{fig:T_C_of_n_Mn}.
The following features are important. First, for all concentrations $x$ 
the value of the Curie temperature
oscillates with occupation. It is negative for $n=-2$ (two additional holes), 
increases with increasing occupation, changes sign and
reaches the maximum in the interval between $n=-1$ (one additional hole) 
and $n=0$. Then it decreases again, changes sign in the interval
between $n=0$ and $n=1$ (one additional electron) reaches minimum and
increases again. The only deviation from this scenario is the case of $x=3.125\%$
where the curve has two maxima, one close to $n=1$ and another to $n=0$.
(The two-maxima shape of the curve can be related to the properties of the 
electron DOS presented in I in the energy region just below 
the Fermi level: the closeness of the top of the
spin-down valence band to the Fermi level and a deep minimum in the spin-up DOS.) 
A remarkable feature 
in the dependences is the kink at  $n=1$ for  $x=3.125\% $ and  $x=6.25\%$. 
This $n$ value corresponds to a completely filled valence band and empty conduction band.
Correspondingly, for  $n<1$ the carriers are holes in the valence band and for
$n>1$ the carriers are electrons in the conduction band. 
The abrupt change of the character
of the carrier states at $n=1$ results in the discontinuity of the 
slope of $T_C(n)$. For $x=12.5\%$ and $x=25\%$
the kink is not obtained because of the overlap of the valence and conduction bands.      
Also the minimum for these concentrations is shifted to a non-integer value of $n$
that depends on the details of the band overlap and cannot be predicted 
without calculations. 

\begin{figure}
\caption{$T_C^{MF}$ of (Ga,Mn)As as a function of the Mn concentration $x$
for different electron numbers $n$. $n$ is defined as in Fig. \ref{fig:T_C_of_n_Mn}.
Note that the number of holes $p$ per Mn atom equals 1-$n$. 
The stars show the experimental values of the
Curie temperature.
\label{fig:gamnas_T_x_n}}
{\includegraphics[width=8cm,angle=-90]{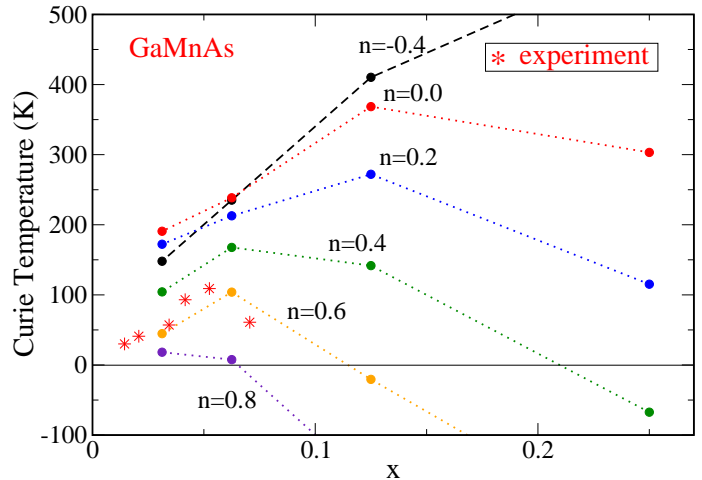}}
\end{figure}

The amplitude of the curves decreases with decreasing $x$. Therefore, potentially
the highest Curie temperature can be reached for the highest Mn concentration.
This, however, needs large-scale tuning of the number of carriers. 
(Both, the growth of the samples with Mn concentration of the order of 20$\%$ 
and corresponding large-scale tuning of the number of carriers are technologically
hardly possible.) It is interesting to consider the dependence of $T_C$ on 
concentration $x$ for various numbers of electrons $n$. 
Since the form
of the curves $T_C(n)$ differs considerably for different $x$
they intersect (Fig. \ref{fig:T_C_of_n_Mn}).
The intersection of the curves 
influences the dependence of 
$T_C$ on concentration $x$ 
(Fig. \ref{fig:gamnas_T_x_n}). Indeed, if before intersection 
at some $n_\circ$, $T_C$ for concentration $x_1$ is larger than  
$T_C$ for concentration $x_2$ after the intersection the relation is opposite.
The last property is reflected in Fig. \ref{fig:gamnas_T_x_n}  
where the value of $T_C^{MF}$ as a function of $x$ is shown for different 
hole concentrations. Indeed, a monotonous dependence for n=$-0.4$ is replaced
by a nonmonotonous behavior for $n=0$ with the maximum at $x=12.5\%$.
For $n=0.4$ and $n=0.6$ the maximum corresponds to $x=6.25\%$. 
The experimental $T_C$ values obtained in Ref. \cite{matsukura}  are in good agreement with 
the values for $n=0.6$ . For $n=0.8$ the maximum value of the Curie temperature
corresponds to the lowest Mn concentration. 

Note that the mean-field approximation usually overestimates the Curie temperature
and the comparison of $T_C^{MF}$ with experiment 
should be made with caution. 
For example, the value of the Curie temperature of (GaMn)As obtained in I within the
random phase approximation (RPA) is about 15-20\% 
smaller than the $T_C^{MF}$. This means that good agreement with the experimental
Curie temperatures just for $n=0.6$ might be accidental. The most important result of the
calculation is the qualitative trends obtained with the variation of the number of
carriers. 

In general, Figs. \ref{fig:T_C_of_n_Mn},\ref{fig:gamnas_T_x_n} confirm the picture of the 
ferromagnetism in DMS as 
mediated by the charge carriers. Indeed, the decrease of the number of holes 
in the valence band ($n>0$) leads to a fast decrease in the trend toward ferromagnetism. 
However, because of the oscillating character of the curves $T_C(n)$, the increase of the 
number of holes to more than one per Mn ($n<0$) has only small potential for obtaining higher
Curie temperature in (GaMn)As. 

Figure \ref{fig:T_C_of_n_Mn} (and the corresponding Figs.
\ref{fig:T_C_of_n_Cr}  and \ref{fig:T_C_of_n_Fe} 
in the following) show that the 
contribution of the completely filled bands into exchange interaction
is always antiferromagnetic. Indeed, in all cases of completely filled bands
($n=1$ in Fig. \ref{fig:T_C_of_n_Mn} , 
$n=-1$ and $n=2$ in Fig.
\ref{fig:T_C_of_n_Cr}  
or $n=2$ in Fig. 
\ref{fig:T_C_of_n_Fe}) 
the estimation of the 
$T_C^{MF}$ is negative. On the other hand, the deviation from the completely
filled bands to smaller $n$ (creating holes in the filled bands) or to larger $n$ 
(creating electrons in the empty bands) leads to increasing $T_C^{MF}$ and, respectively,
to increasing trend to ferromagnetism. 
The results of the study of the two-band model in Sect. \ref{sect:two-band} help to understand 
the physical mechanism of this property in terms of the changes in the band structure with the
deviation of the magnetic configuration from the collinear ferromagnetic one. Indeed,
we have seen that for completely filled bands the changes in the energy
of different states compensate
and no increase of the total energy is connected with the deviation from the collinear
ferromagnetic structure. The antiferromagnetic superexchange obtained in 
Fig.~\ref{fig:T_C_of_n_Mn} for $n=1$ is not described by the model of 
Sect.~\ref{sect:two-band} since the hybridization between magnetic and band 
electrons is not taken into account. For the case there are partially field band
with holes or electron present, the ferromagnetic state becomes energetically preferable.    
These properties correlate with the 
view of the magnetism of the DMS as governed by the 
competition between the antiferromagnetic exchange through completely filled 
bands and ferromagnetic exchange through charge carriers (holes or electrons).

\subsection{(GaCr)As} 
\begin{figure}
\caption{The DOS of Ga$_{1-x}$Cr$_{x}$As. The DOS is given per 
unit cell of the zinc-blende crystal structure. The DOS above(below)
the abscissas axis corresponds to the spin-up(down) states. 
 \label{fig:DOS_GaCrAs}}
{\includegraphics[width=8cm]{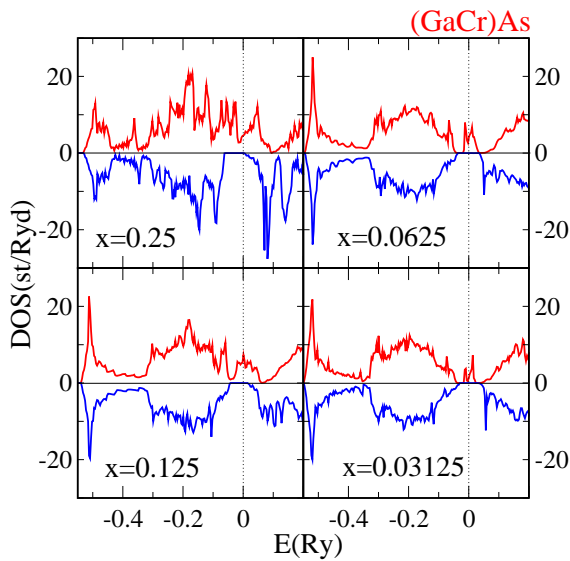}}
\end{figure}
In Fig. \ref{fig:DOS_GaCrAs} we show the DOS of (GaCr)As for four values of the Cr concentration. 
There is substantial difference between these DOS 
and the DOS of (GaMn)As presented in I. In particular, an important different feature is a 
higher energy position of the Cr 3d states
relative to the GaAs states if compared with the position of the Mn 3d states in (GaMn)As. 
For the Cr concentration of $x=3.125\%$ 
this energy shift results in an impurity band lying within the 
semiconducting gap of GaAs 
and separated in energy from both the
valence and the conduction bands. 
(In the case of (GaMn)As corresponding states lie 
at the top of the valence band.)
The impurity band has spin-up character. The replacement
of one Ga atom in the supercell of GaAs by a Cr atom does not change the 
number of the spin-down states in the valence band. 
In the spin-up channel there are, however, five additional
energy bands which are related to the Cr 3d states. Two of them lie
within the valence band and three in the semiconducting energy gap. 
Since there are five extra energy bands 
and only three extra electrons 
(the atomic configurations of Ga and Cr are 4s$^2$4p$^1$ and 3d$^4$4s$^2$) 
the impurity band is not filled.
The integrated number of the occupied (electron) states in the impurity band is 
one per Cr atom (correspondingly, one
electron per supercell). On the other hand, the integrated number of empty (hole) 
states in the impurity band is two per Cr atom (two holes per supercell).

With increasing $x$, the impurity band becomes broader. At $x=6.25\%$ 
it touches the spin-down states of the conduction band still being separated by an 
energy gap from the valence band. At $x=12.5\%$ and $x=25\%$
the impurity band overlaps with both valence and conduction bands. For all 
concentrations the Fermi level lies in the energy gap of the spin-down DOS. 
Therefore for all $x$ the system is half-metallic
and is characterized by a $100\%$ 
spin-polarization of the carriers at the Fermi energy. The latter property is very 
important for efficient spin-injection to semiconductors.

In Table \ref{tab:magnetic_moments} we present 
the magnetic moment of the Cr atom, the induced
moment on the nearest As atom, and the magnetic moment per supercell.
\begin{table}
\caption{Magnetic moments in Ga$_{1-x}$Cr$_{x}$As and Ga$_{1-x}$Fe$_{x}$As. 
There are shown the
moment on the 3d impurity, the induced moment on the nearest As atoms, and the 
magnetic moment of the supercell.
All moments are in units of $\mu_B$. 
\label{tab:magnetic_moments}}
\begin{tabular}{llcccc}
\hline\hline
&&\multicolumn{4}{c}{x}\\
&&0.25&0.125&0.0625&0.03125\\
\hline
        &Cr&3.10 &3.11  &3.13 & 3.15\\
(GaCr)As&As&-0.09 &-0.08  &-0.07 & -0.07\\
      &cell&3.00 &3.00  &3.00 &3.00\\
\\
        &Fe&3.18 &3.29  &3.40 &3.46\\
(GaFe)As&As&0.03 &0.04  &0.05 &0.06\\
      &cell&3.61 &3.87  &4.15 &4.39\\
\hline\hline
\end{tabular}
\end{table} 
Both the Cr and As moments vary weakly with the change of $x$.
This is the result of the half-metalicity that
leads to a fixed number of the spin-up and spin-down electrons in the
system and, correspondingly, complicates the change of atomic moments. 

Taking as an example the system with
$x=3.125\%$ we find that the contribution of the 3d electrons into the 
Cr moment is 3.10$\mu_B$ with the rest 0.05$\mu_B$ coming from the
4s and 4p electrons. The total number of the 3d electrons
in the Cr sphere is $4.38$. The main contribution to the induced As
moment comes from the 4p electrons.  

Interestingly, the induced As moment in (GaCr)As is larger
than in (GaMn)As though the Mn moment is larger than the Cr moment.
The explanation of this fact is in the hybridization of the spin-up states of the 3d 
and As atoms. Because of this hybridization, part of the As spin-up states contributes to the 
hole states in the valence (GaMnAs) or impurity (GaCrAs) band and is unoccupied.
The number of hole states in (GaCr)As is larger than in (GaMn)As, therefore the negative
induced moment on the As atom is larger as well.  

Since the ferromagnetism is expected to be mediated by carriers it is 
important to consider their localization about Cr atoms. 
\begin{figure}
\caption{Distribution of the carrier states over atoms (upper panel) and
induced magnetic moments (lower panel) for  Ga$_{1-x}$Cr$_{x}$As and Ga$_{1-x}$Mn$_{x}$As
with $x=3.125\%$. For (GaCr)As, the distribution is given for the occupied (electron)
states in the impurity band. For (GaMn)As,the distribution is given for empty (hole)
states in the valence band.  
All values are for the coordination spheres. The numbers of atoms in the coordination spheres
are shown at the top of the picture. The calculated values for (GaMn)As are taken from I.
\label{fig:hole_moment_distribution}}
{\includegraphics[width=8.5cm,angle=0]{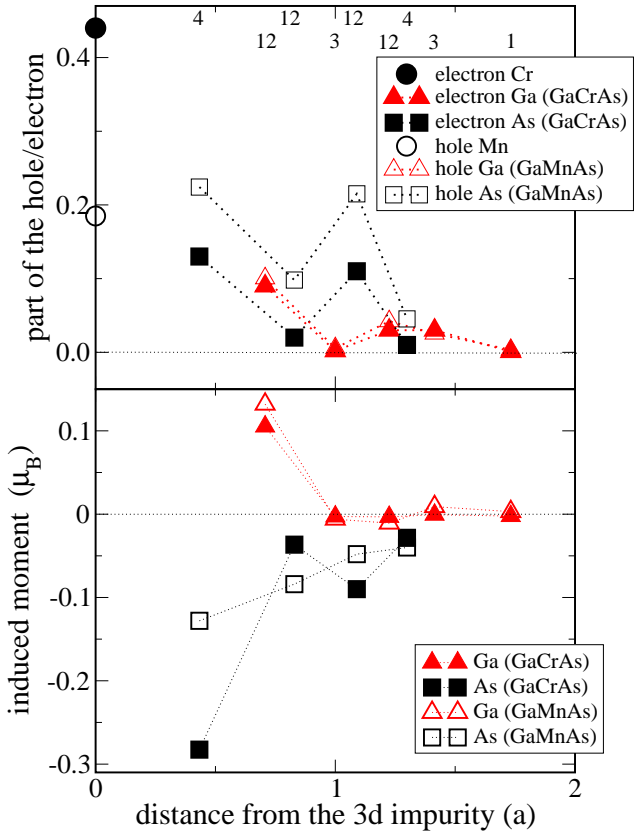}}
\end{figure}
In Fig. \ref{fig:hole_moment_distribution} we show for $x=3.125\%$ 
the distribution in the supercell of the one electron in the occupied part 
of the impurity band. 
This distribution was calculated as follows. At first the distribution over
atoms was found for each state of the occupied part of the impurity band.
Then the summation over these states was performed. The number of electrons
in the occupied part of the impurity band is exactly one per Cr atom
(or, what is equivalent, per supercell). For comparison, we plot in 
Fig.~\ref{fig:hole_moment_distribution} the data for one hole per Mn
atom in (GaMn)As calculated similarly.
About 44\% 
of the electron is in the Cr sphere. The first coordination
sphere of As contains 13\% 
of the state. The third coordination As 
sphere with distance $1.09a$ from the Cr atom contains 11\% of the electron.
We will show that the large portion of the electron states at the atoms of the 
third coordination sphere of the As atom is important for the magnetic
ordering in the system. The first coordination sphere of Ga atoms
contains 9\% 
of the state. 
Each of the other coordination spheres contains less that 3\% of the state.  

Another important characteristic for establishing the exchange interaction between 
the Cr atoms is the value of induced moments (Fig.\ref{fig:hole_moment_distribution}). The induced 
moments on all As atoms are antiparallel to the Cr moment. 
The same feature was obtained in I for (GaMn)As (see Fig.\ref{fig:hole_moment_distribution}) 
and is explained by the
contribution of the As spin-up states to the hole states in the impurity band
discussed above. On the Ga atoms, the induced moments are parallel to the Cr moment. 
Note that the parts of the states corresponding to the Ga atoms and the induced 
moments on the Ga sites are very similar in both (GaCr)As and (GaMn)As. On the other
hand, for the 3d and As atoms there is substantial difference between two systems. 

For (GaCr)As, there is a clear correlation
between the value of the induced moment and the part of the electron states
in the impurity band corresponding to a given atom (Fig.\ref{fig:hole_moment_distribution}). 
In particular there is a sizable 
induced moment at the third coordination sphere of As atoms. Note that 
the hole states in (GaMn)As are less localized about Mn atom compared with the electron
states localization about Cr atoms in (GaCr)As. 
On the other hand, the values of the induced moments are 
larger in (GaCr)As. Thus apriori it is not clear which of the two systems 
has potential for the ferromagnetism with higher Curie temperature.

\begin{figure}
\caption{The parameters of the exchange interaction between Cr atoms (upper panel)
and the variation of $T_C^{MF}$ with increasing number of the contributing 
coordination spheres (lower panel). The abscissa gives the radius of the coordination sphere
in the units of the lattice parameter of the zinc-blende crystal structure. The calculations
has been performed up to $r=7a$. Because of very small values the exchange parameters for
large $r$ are not shown. The broken vertical lines show the coordination spheres with 
strongest exchange interaction with the central atom.
Two data points for the same system and the same radius (e.g., at $r=2.828a$
for $x=12.5\%$ and ) mean that
there are two inequivalent coordination spheres with equal radii. 
\label{fig:exch_param_cr}}
{\includegraphics[width=8cm]{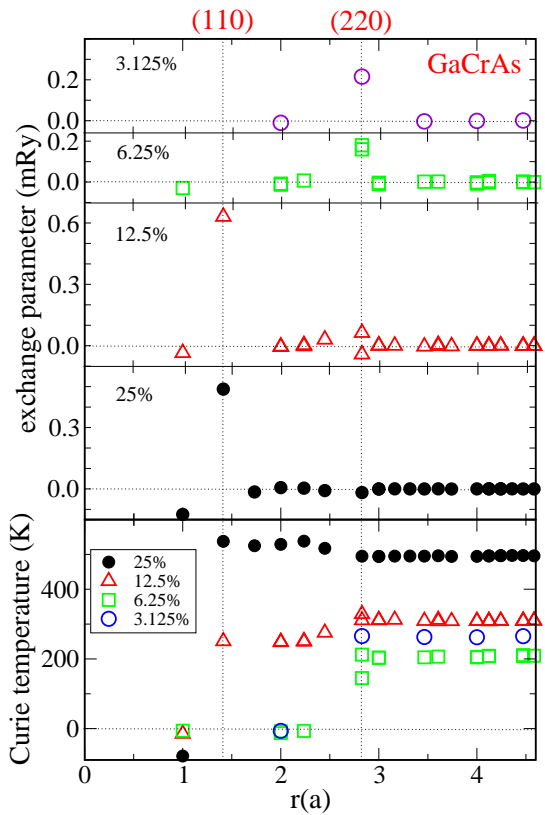}} 
\end{figure}

\begin{figure}
\caption{The absolute values of the exchange parameters between 3d impurities
as a function of the interatomic distance. The impurity concentration is 25\%.
The [110] crystallographic direction is presented. 
A logarithmic ordinate axis is used to visualize fast decay of the exchange
parameters. The small value of the exchange parameter of (GaFe)As at 
$r=2\sqrt 2a$ results from the oscillation of the parameter as a function
of $r$(Fig. \ref{fig:exch_param_fe}).  
The dashed straight line is a guide for the eye.
\label{fig_exch_log}}
{\includegraphics[width=8cm,angle=-90]{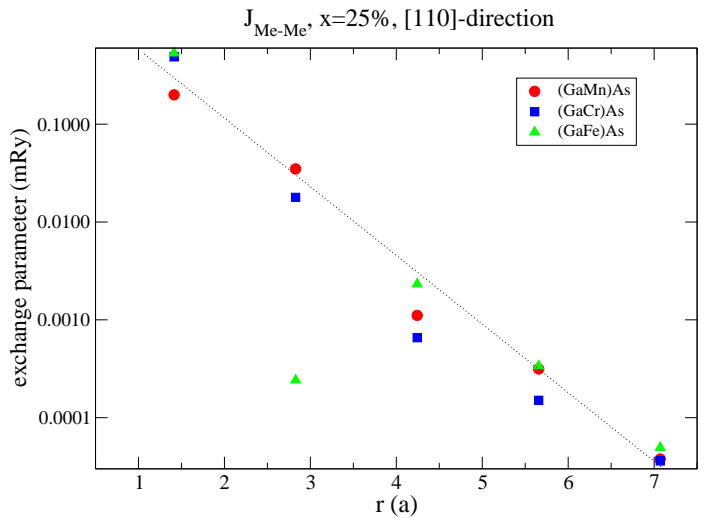}} 
\end{figure}

In Fig. \ref{fig:exch_param_cr} we show calculated interatomic exchange parameters. 
The quantitative comparison 
with the exchange parameters of (GaMn)As presented in I shows that they are substantially different.
There is however a number of qualitative similarities. First, in both cases the Heisenberg 
model with the interaction between the first nearest neighbors only
is not able to describe the magnetism of the system.
Second, the exchange interactions are rather quickly
decreasing with increasing distance between atoms.  
In Fig. \ref{fig_exch_log} we plot the dependence of the 
absolute value of the exchange parameters for three different impurities
on the interatomic distance using logarithmic ordinate axis. 
The decrease of exchange parameters is close to exponential
that is expected for semiconducting and half-metallic systems.\cite{pakutu,kudrn}  
In Fig. \ref{fig:exch_param_cr} we show the variation of the 
Curie temperature [Eq. (\ref{eq:Tc_MF})] 
with increasing number of the contributing coordination spheres. 
For instance, no noticeable contribution to the
Curie temperature is obtained from the interactions between Mn atoms at the
distances larger than 3$a$.     
Third, the dependence of the exchange parameters on the distance between 
Cr atoms is not monotonous. Forth, the coordination spheres which contribute
importantly into the magnetism of (GaMn)As provide, in most cases, 
such contribution also in (GaCr)As. 

\begin{figure}
\caption{$T_C^{MF}$ of (Ga,Cr)As, (Ga,Mn)As, and (Ga,Fe)As for 
different concentrations of the 3d impurity. 
\label{fig:T_C_of_x_GaMeAs}}
\centerline
{\includegraphics[width=7.5cm,angle=-90]{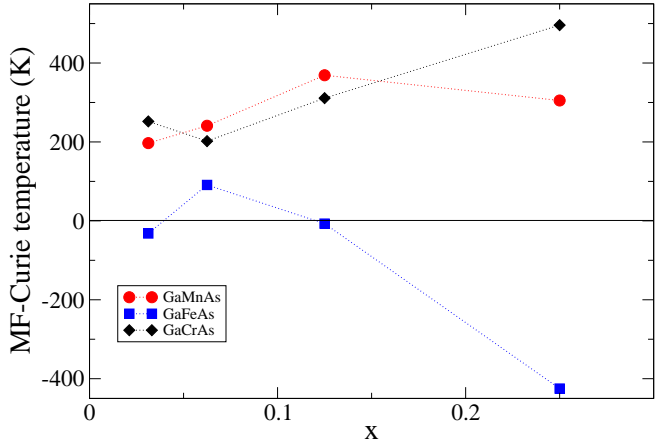}}
\end{figure}
In Fig. \ref{fig:T_C_of_x_GaMeAs} the Curie temperature of (GaCr)As for four values of the 
Cr concentration is presented. For $x=3.125\%$ and $x=25\%$ 
the Curie temperature
is higher than in (GaMn)As, for $x=6.25\%$ and $x=12.5\%$ the relation is opposite.
In general both systems give $T_C$ in the same temperature interval.
Only for very large $x$ the Curie temperature of (GaCr)As becomes
substantially larger. 

The calculated exchange parameters combined with the information about the
induced moments and spatial distribution of the states of the impurity band
allow to suggest the exchange path responsible for the realization of the 
ferromagnetic state. 

The analysis of the interatomic exchange interactions and their contribution
into the Curie temperature (Fig. \ref{fig:exch_param_cr}) shows that the main role
for all concentrations of the Cr atoms is played by the exchange interaction
between the Cr atoms separated by the vectors parallel to (1,1,0) 
(or to the vectors crystallographically equivalent to (1,1,0)). Indeed, for
concentrations $x=25\%,12.5\%$ the main contribution to the Curie temperature
comes from the atoms separated by (1,1,0). For lower concentrations
the main role is played by the interaction between atoms separated by (2,2,0).  

\begin{figure}
\caption{The $z=0$ and $z=\frac{a}{4}$ planes of the supercell for the concentration
of magnetic impurity of 3.125\%. 
The exchange path shown by the thick line connecting the impurity atoms goes
over As atoms.
\label{fig:exch_path}}
\centerline
{\includegraphics[width=8cm,angle=-90]{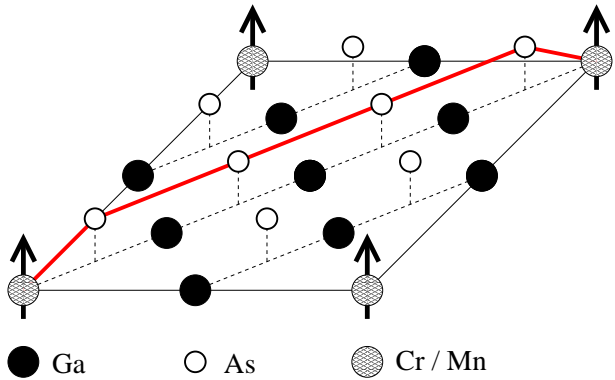}} 
\end{figure}

This result together with the information about the induced moments
and the localization of the carrier states allow to propose the exchange path
for the realization of the strong exchange interactions.  
Let us consider again the case of $x=3.125\%$ (Fig. \ref{fig:exch_path}). 
On the straight line connecting
the Cr atoms at (0,0,0) and (2,2,0) there are three Ga atoms at 
$(\frac{1}{2},\frac{1}{2},0)$, $(1,1,0)$, and $(\frac{3}{2},\frac{3}{2},0)$.
(The first and the third atoms are equivalent.) Since, according to Fig. \ref{fig:exch_param_cr},
only very small portion of the carrier states corresponds to these atoms and the induced 
moments on these atoms are also very small they cannot efficiently mediate the
exchange interaction between Cr atoms. 

On the other hand a substantial part of the electron states in the impurity band is on the
As atoms at $(\frac{1}{4},\frac{1}{4},\frac{1}{4})$ and $(\frac{3}{4},\frac{3}{4},\frac{1}{4})$. 
Simultaneously, these atoms carry sizable induced moments. On the basis of this information we suggest
the following exchange path: Cr(0,0,0)$-$As$(\frac{1}{4},\frac{1}{4},\frac{1}{4})-
{\rm As}(\frac{3}{4},\frac{3}{4},\frac{1}{4})$$-$As$(\frac{5}{4},\frac{5}{4},\frac{1}{4})$ 
$-$As$(\frac{7}{4},\frac{7}{4},\frac{1}{4})$$-$Cr(2,2,0).
Here As$(\frac{7}{4},\frac{7}{4},\frac{1}{4})$ is equivalent to As$(\frac{1}{4},\frac{1}{4},\frac{1}{4})$
and As$(\frac{3}{4},\frac{3}{4},\frac{1}{4})$ is equivalent to As$(\frac{5}{4},\frac{5}{4},\frac{1}{4})$.
The calculational information for (GaMn)As (see I) allows to suggest the same exchange path 
also for this system.

\begin{figure}
\caption{$T_C^{MF}$ of (Ga,Cr)As with different Cr concentrations
as a function of the electron number $n$.
$n=0$ corresponds to the system  Ga$_{1-x}$Cr$_{x}$As with no
additional donor or acceptor defects.  
\label{fig:T_C_of_n_Cr}}
{\includegraphics[width=7.5cm,angle=-90]{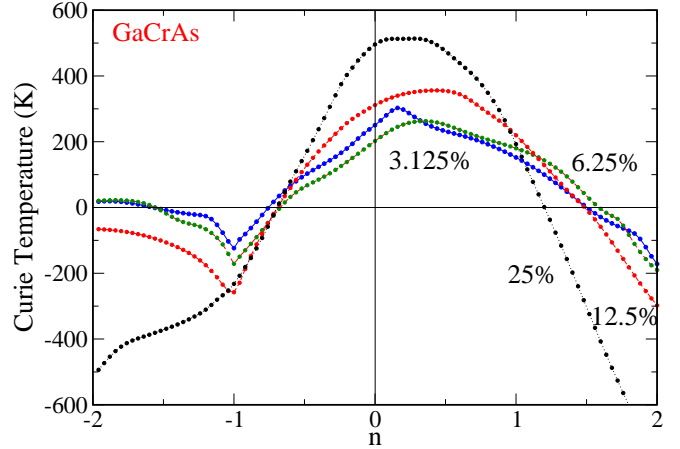}}
\end{figure}

To get deeper insight into the magnetism of (GaCr)As we calculated
the dependence of the Curie temperature on the electron occupation (Fig. \ref{fig:T_C_of_n_Cr}).
Many qualitative features of these dependences are similar to those for 
(GaMn)As (Fig. \ref{fig:T_C_of_n_Mn}). The dependences show oscillating behavior with the 
amplitude decreasing with decreasing Cr concentration. For $x=3.125,6.25,12.5\%$
there is a kink at $n=-1$ similar to the kink at $n=1$ for (GaMn)As. The origin
of the kink is also similar to that for (GaMn)As and consists in the transition 
from the occupation
of the valence-band states at $n<-1$ to the occupation of the impurity-band states
at $n>-1$. At $n=-1$ the valence band is completely filled and the impurity band
is empty. The curves for $x=3.125\%$ and $x=6.25\%$ have another kink at $n=2$ which marks
the end of the filling of the impurity band and beginning of the filling
of the conduction band. 
The reason for the increase of the interatomic ferromagnetic interactions 
with appearance of holes in valence band or electrons in conduction band
was discussed in Sect.~\ref{sect:two-band} in terms of two-band model
and consists in the increase of the energy of the system with deviation
of the moments from the parallel directions. 

An important conclusion from the comparison of the
Figs. \ref{fig:T_C_of_n_Cr}  and \ref{fig:T_C_of_n_Mn} is that the role played in the case of 
(GaMn)As by holes in
the valence band is played in (GaCr)As by the electrons in the impurity band.
In particular, the decrease of the number of electrons results in fast 
decrease of the Curie temperature similar to the behavior obtained in (GaMn)As with
decreasing number of holes. On the other hand, the decrease of the number 
of holes in (GaCr)As (the region of positive $n$ in Fig.\ref{fig:T_C_of_n_Cr}) 
first leads to a moderate increase of the Curie temperature. Only for larger $n$
($n>1$ for $x=3.125,6.25,12.5\%$) a strong decrease of $T_C$ is observed. (Again, a
similar behavior is obtained for (GaMn)As but with $n$ decreasing lower than $-1$.)
Therefore the Curie temperature in (GaCr)As
is less sensitive to the presence of the As antisites and other nonmagnetic donor defects.
Since the formation of the As antisites is difficult to avoid this property is of
technological importance.

\subsection{(GaFe)As}

\begin{figure}
\caption{The DOS of Ga$_{1-x}$Fe$_{x}$As. The DOS is given per 
unit cell of the zinc-blende crystal structure. The DOS above(below)
the abscissas axis corresponds to the spin-up(down) states. 
 \label{fig:DOS_GaFeAs}}
{\includegraphics[width=8cm,angle=-90]{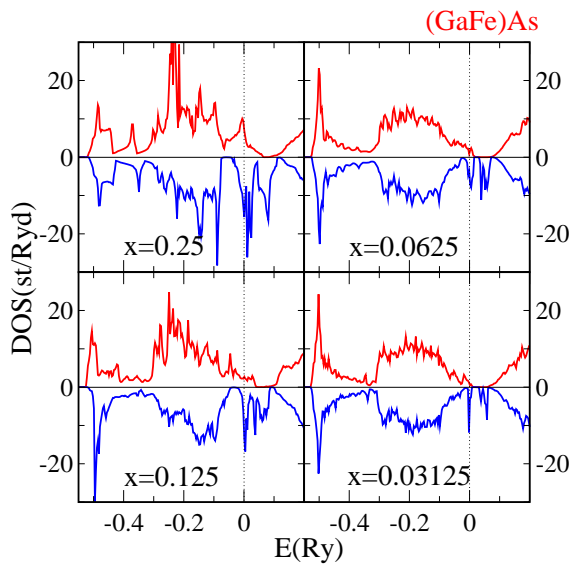}}
\end{figure}

An Fe atom has one additional valence electron compared with a Mn atom. 
Thus no hole or electron carriers
can be expected  in (GaFe)As on the basis of the electron count.
The system can be expected to be strongly 
antiferromagnetic. However, 
the calculations show that the electron structure of 
(GaFe)As differs strongly from the electron structure 
of both (GaMn)As and (GaCr)As and a simple consideration on the basis of the 
electron count does not apply. 
The origin of this difference is in the lower energy position of the
Fe 3d states with respect to the GaAs states if compared with the position 
of the Mn states (Fig. \ref{fig:DOS_GaFeAs}). 
As a result, the spin-down impurity states lie at the top of the 
valence band and become partly occupied. 
Correspondingly, the spin-up impurity states at the top
of the valence band are not completely filled.
The presence of both spin-up and spin-down states at the Fermi level is 
disadvantageous for the realization of an efficient spin-injection
that needs very high spin-polarization of the
states at the Fermi level. 
With increasing $x$ the impurity bands broaden. 
For none of the concentrations the state of the system is half-metallic.

In Table \ref{tab:magnetic_moments} the Fe magnetic moment as well as the induced moment 
on the first As 
atom are shown. Compared with (GaCr)As, the moments strongly vary with concentration. This
variation is a consequence of the presence of both spin-up and spin-down states 
at the Fermi level that leads to strong electron transfer between two spin channels.
The induced moment is now positive and its value increases with decreasing $x$.
The reason for the change of the sign of the induced As moment is directly related to 
the increased number of 3d electrons in an Fe atom compared to Mn and Cr atoms. 
Indeed, in (GaFe)As the number of the hole spin-up states becomes very small.
These hole states play a crucial role in the formation of the negative induced
moments in (GaMn)As and (GaCr)As
since they decrease the number of the As spin-up electron. Simultaneously, 
the number of the As spin-down electrons in (GaFe)As decreases since, because of 
a lower energy position
of the Fe 3d spin-down states, the As spin-down states hybridize with
empty impurity states. 

\begin{figure}
\caption{The parameters of the exchange interaction between Fe atoms (upper panel)
and the variation of $T_C^{MF}$ with increasing number of the contributing 
coordination spheres (lower panel). The abscissa gives the radius of the coordination sphere
in the units of the lattice parameter of the zinc-blende crystal structure. 
The decay of the exchange parameters is illustrated in the logarithmic scale in 
Fig.~\ref{fig_exch_log}. 
\label{fig:exch_param_fe}} 
{\includegraphics[width=10cm]{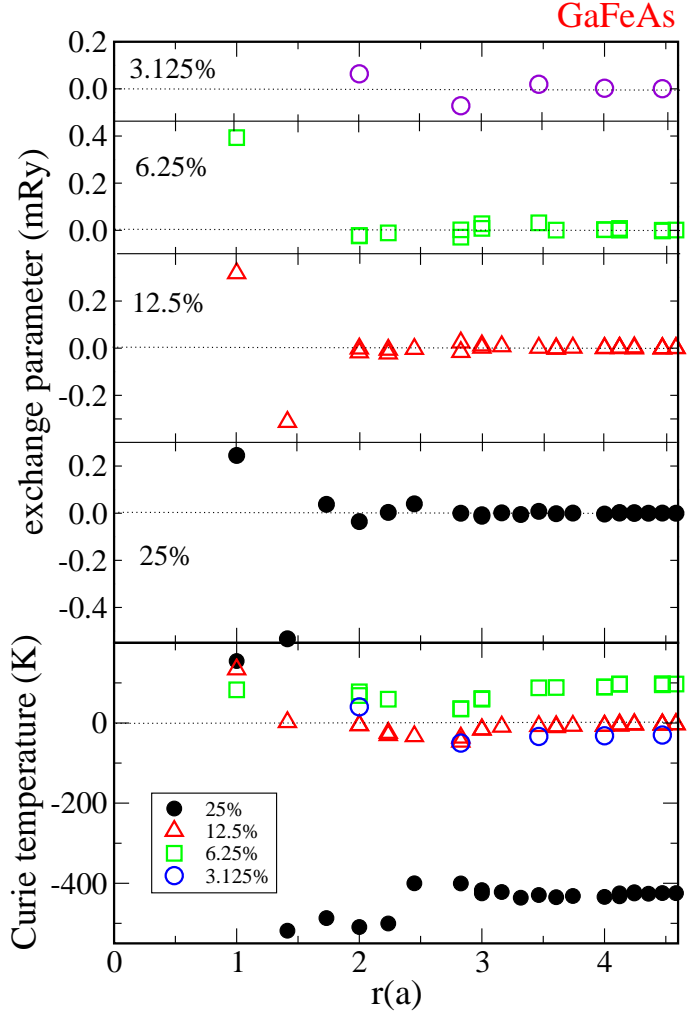}} 
\end{figure}

\begin{figure}
\caption{$T_C^{MF}$ of (Ga,Fe)As with different Fe concentrations
as a function of the electron number $n$.
$n=0$ corresponds to the system  Ga$_{1-x}$Fe$_{x}$As with no
additional donor or acceptor defects.  
\label{fig:T_C_of_n_Fe}}
{\includegraphics[width=7.5cm,angle=-90]{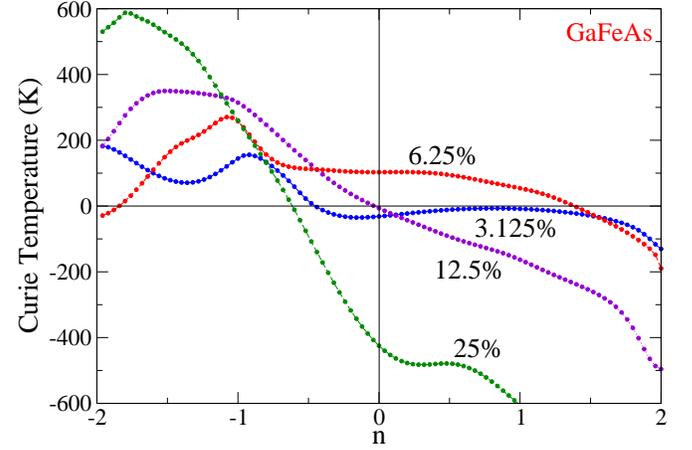}}
\end{figure}

The calculation of the exchange interactions shows (Fig. \ref{fig:exch_param_fe}) 
that they are much more
antiferromagnetic than for (GaMn)As and (GaCr)As. 
Correspondingly, the estimation
of the Curie energy (Fig. \ref{fig:T_C_of_x_GaMeAs}) is negative for $x=3.125\%$ and $x=25\%$  and 
close to zero for $x=12.5\%$. 
It is positive for $x=6.25\%$ 
but much smaller than the Curie temperature for the same concentration of the 3d element 
in (GaMn)As and (GaFe)As. The explanation for such behavior is in small number
of the spin-up holes. The main number of the states at 
and above the Fermi level are of the spin-down type. The spin down hole
states are strongly localized on the Fe atom (76.5\%) 
and do not mediate 
ferromagnetic interactions efficiently. The curves of $T_C(n)$ (Fig. \ref{fig:T_C_of_n_Fe}) 
show that there is a general trend to further strengthening of the antiferromagnetic
interactions with increasing $n$. 
In this case the valence band is filled at $n=2$.
In agreement with the two-band model of Sect.~\ref{sect:two-band} the 
curves have minimum at this point. Appearance of holes ($n<2$) leads to
increasing ferromagnetic interactions.
Although strong acceptor doping has some potential
for increasing ferromagnetic interactions, (GaFe)As  is not a good candidate 
for the application in the semiconductor spin-electronic devices.  
 
\section{Conclusions}

In the given paper 
we complete our study of the electronic structure, exchange interactions and Curie
temperature
of (GaMn)As started in I and report calculations for
(GaCr)As and (GaFe)As. A different number of the valence electrons in Cr, Mn and Fe
allows the investigation of the trends within the series. We find that there is a
similarity in the character of the exchange interactions of 
(GaMn)As and (GaCr)As. 
The dependence of the exchange interactions and Curie temperature
on the band occupation shows that the presence of the charge carriers is crucial
for establishing of the ferromagnetism in (GaMn)As and (GaCr)As. 
We find that the role played in (GaMn)As by the holes
in the valence band is played in (GaCr)As by the electrons in the impurity band.
An important difference between two systems is in the 
character of the dependence on the variation of the number of carriers. The 
ferromagnetism in (GaMn)As is very sensitive to the presence of the donor defects,
like As$_{\rm Ga}$ antisites. On the other hand, the Curie temperature of (GaCr)As
depends rather weakly on the presence of this type of defects but decreases strongly
with decreasing number of electrons.
The properties of (GaFe)As are
found to differ
crucially from the properties of (GaCr)As and (GaMn)As. (GaFe)As does not show a trend to
ferromagnetism and is not half-metallic that makes this system unsuitable for 
the use in spintronic semiconductor devices.   

We show that the sign of the induced moment on the As atoms and, correspondingly,
the sign of the p-d exchange (often denoted as $N\beta$) varies from negative in 
(GaMn)As and (GaCr)As to positive in (GaFe)As. There is a connection between the half-metalicity of the 
system and the negative sign of $N\beta$. 

We show that the strongest exchange interactions in (GaCr)As and (GaMn)As
are between the 3d atoms separated by the vector parallel to (1,1,0) or
to a vector crystallographically equivalent to (1,1,0). On the basis of the
calculated results we propose the exchange path over As atoms that is 
responsible for the mediating the ferromagnetic interactions between 
the Cr atoms.
 
\begin{acknowledgments}
The financial support of Bundesministerium f\"ur Bildung und
Forschung is acknowledged. 
\end{acknowledgments}



\begin{thebibliography}{99}
\bibitem{ohno_gaas}  H. Ohno, A. Shen, F. Matsukura, A. Oiwa, A. Endo,
S. Kutsumoto,and Y. Iye, Appl. Phys. Lett. {\bf 69}, 363 (1996).
\bibitem{dietl_rev}
T. Dietl, Semicond. Sci. Technol. {\bf 17}, 377 (2002).
\bibitem{dms_mac}
J. K\"onig, J. Schliemann, T. Jungwirth, and A.H. MacDonald, 
cond-mat/0111314 (unpublished).
\bibitem{sanvito_rev}
S. Sanvito, G. J. Theurich and N. A. Hill,
J. Superconductivity 15, 85 (2002). 
\bibitem{diofma}
T. Dietl, H. Ohno, and F. Matsukura, Phys. Rev. B {\bf 63} , 195205 (2001).
\bibitem{mark01}
 M. van Schilfgaarde, and O. N. Mryasov, Phys. Rev. B {\bf 63} , 233205 (2001).
\bibitem{bopa}
G. Bouzerar and T. P. Pareek, Phys. Rev. B {\bf 65}, 153203 (2002);
G. Bouzerar, J. Kudrnovsky, and P. Bruno, cond-mat/0208596.
\bibitem{jukosi}
T. Jungwirth, J. K\"onig, J. Sinova, J. Kucera, and A. H. MacDonald,
Phys. Rev. B {\bf 66}, 012402 (2002).  
\bibitem{erpe}
S. C. Erwin and A. G. Petukhov, Phys. Rev. Lett. {\bf 22}, 227201 (2002).
\bibitem{sabr02_1}
L. M. Sandratskii and P. Bruno, Phys. Rev. B {\bf 66}, 134435 (2002).
\bibitem{sato_rev}
K. Sato, and H. Katayama-Yosida, Semicond. Sci. Technol. {\bf 17}, 367 (2002).
\bibitem{sazaak}
R. Moriya, Y. Katsumata, Y. Takatani,
S. Haneda, T. Kondo, and H. Munekata, Physica E {\bf 10}, 224 (2001). 
\bibitem{mokata}
H. Saito, W. Zaets, R. Akimoto, K. Ando, Y. Mishima, and M. Tanaka, J. Appl. Phys. {\bf 89}, 
7392 (2001).
\bibitem{akmash} 
H. Akinaga, T. Manago, and M. Shirai, Jpn. J. Appl. Phys., {\bf 39}, L1118 (2000).
\bibitem{akai}
 H. Akai, Phys. Rev. Lett. {\bf 81}, 3002 (1998).
\bibitem{magnetism_anderson} 
P. W. Anderson and H. Hasegawa, Phys. Rev. {\bf 100}, 675 (1955);
P. W. Anderson, in {\it Magnetism I}, edited by Rado and Suhl
(Academic Press, New York and London, 1963).
\bibitem{sand86}
L. M. Sandratskii, Phys. Stat. Solidi (b) {\bf 135}, 167 (1986).
\bibitem{adv}
L. M. Sandratskii, Advances in Physics {\bf 47}, 91 (1998).
\bibitem{deGennes}
P.-G. de Gennes, Phys. Rev. {\bf 118}, 141 (1960).
\bibitem{mrsafr}
O. N. Mryasov, R. F. Sabiryanov, A. J. Freeman, and S. S. Jaswal,
Phys. Rev. B {\bf 56}, 7255 (1997).
\bibitem{comm2} Note, that in Refs. \cite{magnetism_anderson,deGennes}
the notion of ''double exchange'' was used to study the ferromagnetic exchange
interaction in mixed-valence systems. On the other hand, the same notion was used 
in Refs. \cite{mrsafr,akai,mark01} for the interpretaion of the DFT calculations
where no mixed valency was present. The use of this notion is justified by the fact
that the DFT calculations capture the same physics that is cosidered in the
original definition of the double exchange: the lowest kinetic energy of the 
charge carriers for the ferromagnetic configuration of the strong local exchange fields.
\bibitem{wikuge}
A. R. Williams, J. K\"ubler, and C. D. Gelatt, 
Phys. Rev. B {\bf 19}, 6094(1979). 
\bibitem{ohno_jmmm}
H. Ohno, J. Magn. Magn. Mater. {\bf 200}, 110 (1999).
\bibitem{sanv_apl01}
S. Sanvito, and N. A. Hill, Appl. Phys. Lett. {\bf 78}, 3493 (2001).  
\bibitem{korz02}
P. A. Korzhavyi, I. A. Abrikosov, E. A. Smirnova, L. Bergqvist, P. Mohn, R. Mathieu, P. Svedlindh, 
J. Sadowski, E. I. Isaev, Yu. Kh. Vekilov, and O. Eriksson,
Phys. Rev. Lett. {\bf 88}, 187202 (2002).
\bibitem{comm1} 
The data for x=6.25$\%$ was presented in I
\bibitem{matsukura}
F. Matsukura, H. Ohno, A. Shen, and Y. Sugawara,
Phys. Rev. B {\bf 57}, 2037 (1998).
\bibitem{pakutu}
M. Pajda, J. Kudrnovsky, I. Turek, V. Drchal, and P. Bruno,
Phys. Rev. B {\bf 64}, 174402 (2001).
\bibitem{kudrn}
An exponential decrease of the exchange parameters of half-metallic DMS
was also obtained by J. Kudrnovsky (private communication) 
in the calculation within the coherent-potential approximation. 
\end{thebibliography}
\end{document}